\documentclass[doublecol]{epl2}
\usepackage{amsmath}
\usepackage{amssymb}
\usepackage{graphics}
\usepackage{epsfig}

\title{Superconductivity and physical properties of strongly electron correlated compounds La$_{n}$Ru$_{3n-1}$B$_{2n}$ ($n=$ 1, 2, and 3)}
\shorttitle{Superconductivity and physical properties of La$_{n}$Ru$_{3n-1}$B$_{2n}$}
\author{Hechang Lei\thanks{E-mail: \email{hlei@lucid.msl.titech.ac.jp}} and H. Hosono}
\shortauthor{H. C. Lei and H. Hosono}
\institute{Frontier Research Center, Tokyo Institute of Technology, Yokohama 226-8503, Japan}

\abstract{
We examined the physical properties of homologous La$_{n}$Ru$_{3n-1}$B$_{2n}$ ($n=1-3$) series including a new compound of $n=$ 2. All of these compounds showed strong electron-electron correlation characterized by large Wilson ratio. In contrast to LaRu$_{2}$B$_{2}$ and La$_{2}$Ru$_{5}$B$_{4}$ that show normal metal behaviors down to 1.8 K, we discover La$_{3}$Ru$_{8}$B$_{6}$ is an intermediately coupled BCS superconductor with $T_{c}\sim$ 3.2 K. The experimental and theoretical calculation results suggest that the emergence of superconductivity in La$_{3}$Ru$_{8}$B$_{6}$ attributes to the rather large density of states at $E_{F}$ when compared to other two compounds.
}


\begin{document}

\maketitle

\section{Introduction}

Ternary boride superconductors with VIIIB precious metals (Ru, Rh, Pd, Os, Ir, Pt) have been studied for a long time because of their various exotic properties. For example, noncentrosymmetric Li$_{2}$Pt$_{3}$B with strong antisymmetric spin-orbit coupling exhibits the mixing of spin-singlet and spin-triplet pairings, when compared to the BCS $s$-wave superconductor Li$_{2}$Pd$_{3}$B \cite{Togano,Badica,Yuan}. Another series compounds RERh$_{4}$B$_{4}$ (RE = rare earth elements) show the coexistence of long-range ferromagnetic (FM) order and superconductivity, which has been extensively studied over the past 30 years \cite{Matthias,Maple}.

Recently, superconductivity is discovered in A$_{3}$Rh$_{8}$B$_{6}$ (A = Ca, Sr) compounds \cite{Takeya1}. They belong to a homologous series A$_{n}$Rh$_{3n-1}$B$_{2n}$ \cite{Jung}. The measurements of physical properties indicate that Ca$_{3}$Rh$_{8}$B$_{6}$ and Sr$_{3}$Rh$_{8}$B$_{6}$ are superconductors with $T_{c}=$ 4.0 and 3.4 K, respectively, however, AM$_{2}$B$_{2}$ ($n=$ 1, M = Rh, Ir) do not show superconductivity for $T>$ 1.8 K \cite{Takeya1}. The experimental and theoretical calculation results suggest that the main reason of the emergence of superconductivity in A$_{3}$Rh$_{8}$B$_{6}$ should be due to the large $N(E_{F})$ (density of states (DOS) at fermi energy level, $E_{F}$) \cite{Takeya1}. On other hand, isostructural compound Y$_{3}$Os$_{8}$B$_{6}$ also shows superconductivity with $T_{c}=$ 5.8 K \cite{Lopes}. Therefore, it is interesting to study the physical properties of other isostructural compounds.

We notice that the similar homologous series exists in La$_{n}$Ru$_{3n-1}$B$_{2n}$ ($n=$ 1, 3, and $\infty$). Among them, only the physical properties of one end member La$_{1-\delta}$Ru$_{3}$B$_{2}$ ($\delta \sim$ 0.1) has been studied and it does not show superconductivity down to 1.2 K \cite{Ku}. In this work, we study the physical properties of La$_{n}$Ru$_{3n-1}$B$_{2n}$ ($n=$ 1 - 3). The superconductivity is found in La$_{3}$Ru$_{8}$B$_{6}$ with $T_{c}\sim$ 3.2 K and the superconducting properties is characterized in detail.

\section{Experiment}

Polycrystal samples of La$_{n}$Ru$_{3n-1}$B$_{2n}$ were synthesized by arc-melting method. The stoichiometric La, Ru and B elements are arc-melted under high-purity Ar on a water-cooled copper hearth. For homogeneity, samples were remelted several times. The losses of mass during arc-melting process are less than 1\%. The powder x-ray diffraction (PXRD) patterns were collected by using a Bruker diffractometer model D8 ADVANCE (Cu rotating anode). Rietveld refinement of the XRD patterns was performed using the code TOPAS4.2 \cite{topas}. The resistivity $\rho(T)$ was measured using a four-probe configuration on rectangular bar of polycrystalls. Heat capacity was measured by thermal relaxation method. Electrical transport and heat capacity measurements were carried out in PPMS-9 from 1.9 to 300 K. Magnetization measurements were performed using a Quantum Design vibrating sample magnetometer from 1.8 to 300 K. First principle electronic structure calculations were performed using experimental crystallographic parameters within the full-potential linearized augmented plane wave (LAPW) method \cite{Weinert} implemented in the WIEN2k package \cite{Blaha}. The general gradient approximation (GGA) of Perdew \textit{et al.} \cite{Perdew}, was used for exchange-correlation potential. The product of the muffin tin radius ($R_{MT}$) and the largest wave number of the basis set ($K_{max}$) is fixed at 7.0 for all of calculations. The $R_{MT-La}$ is 1.32 \AA\ for all of three compounds. The $R_{MT-Ru}$ and $R_{MT-B}$ are 1.14/0.84, 1.22/0.90 and 1.15/0.85 \AA\ for LaRu$_{2}$B$_{2}$, La$_{2}$Ru$_{5}$B$_{4}$ and La$_{3}$Ru$_{8}$B$_{6}$, respectively. Self-consistency was carried out on 1000 k-point meshes in the whole Brillouin zone.

\section{Results and Discussion}

\begin{figure}[tbp]
\centerline{\includegraphics[scale=0.38]{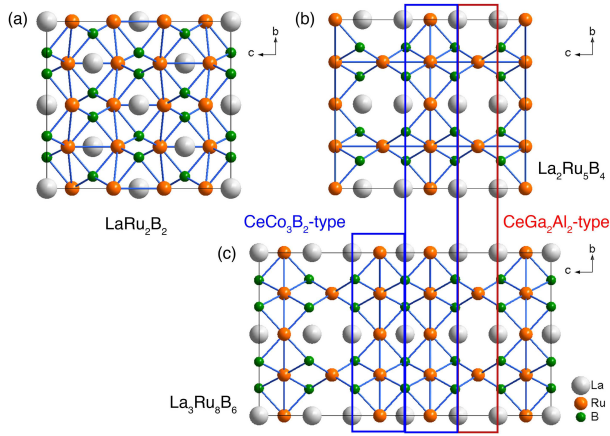}} \vspace*{-0.3cm}
\caption{Crystal structures of (a) LaRu$_{2}$B$_{2}$, (b) La$_{2}$Ru$_{5}$B$_{4}$, and (c) La$_{3}$Ru$_{8}$B$_{6}$. The CeCo$_{3}$B$_{2}$-type and CeGa$_{2}$Al$_{2}$-type structural fragments are highlighted by blue and red rectangles.}
\end{figure}

The structure of LaRu$_{2}$B$_{2}$ (fig. 1(a)) is closely related to CeCo$_{3}$B$_{2}$-type structure. For LaRu$_{2}$B$_{2}$, 1/3 of Ru atoms are removed in the puckered Ru Kogam\'{e} layers when compared to CeCo$_{3}$B$_{2}$-type compounds, forming Ru chains in $ab$-plane. Along the $c$-axial direction, the layers are shifted $b/2$ from each other \cite{Horvath}. Moreover, in contrast to CaRh$_{2}$B$_{2}$ with space group Fddd, the rather small symmetry deviations in LaRu$_{2}$B$_{2}$ lower the space group to the noncentrosymmetric crystallographic subgroup F222 \cite{Horvath}. LaRu$_{2}$B$_{2}$ can also be viewed as the stacking variants of CeAl$_{2}$Ga$_{2}$ type \cite{Sologub1}. On the other hand, the structures of La$_{2}$Ru$_{5}$B$_{4}$ and La$_{3}$Ru$_{8}$B$_{6}$ as shown in fig. 1(b) and (c) represent the intergrowth of CeAl$_{2}$Ga$_{2}$-type LaRu$_{2}$B$_{2}$ and CeCo$_{3}$B$_{2}$-type LaRu$_{3}$B$_{2}$ structural fragments along $c$ axis, taken in ration 1 : 1 and 1 : 2, respectively \cite{Sologub1}. The formula for the series can be expressed by  La$_{n}$Ru$_{3n-1}$B$_{2n}$ = LaRu$_{2}$B$_{2}$ + $(n-1)$LaRu$_{3}$B$_{2}$. When $n$ goes infinity, we will get LaRu$_{3}$B$_{2}$.

\begin{figure}[tbp]
\centerline{\includegraphics[scale=0.6]{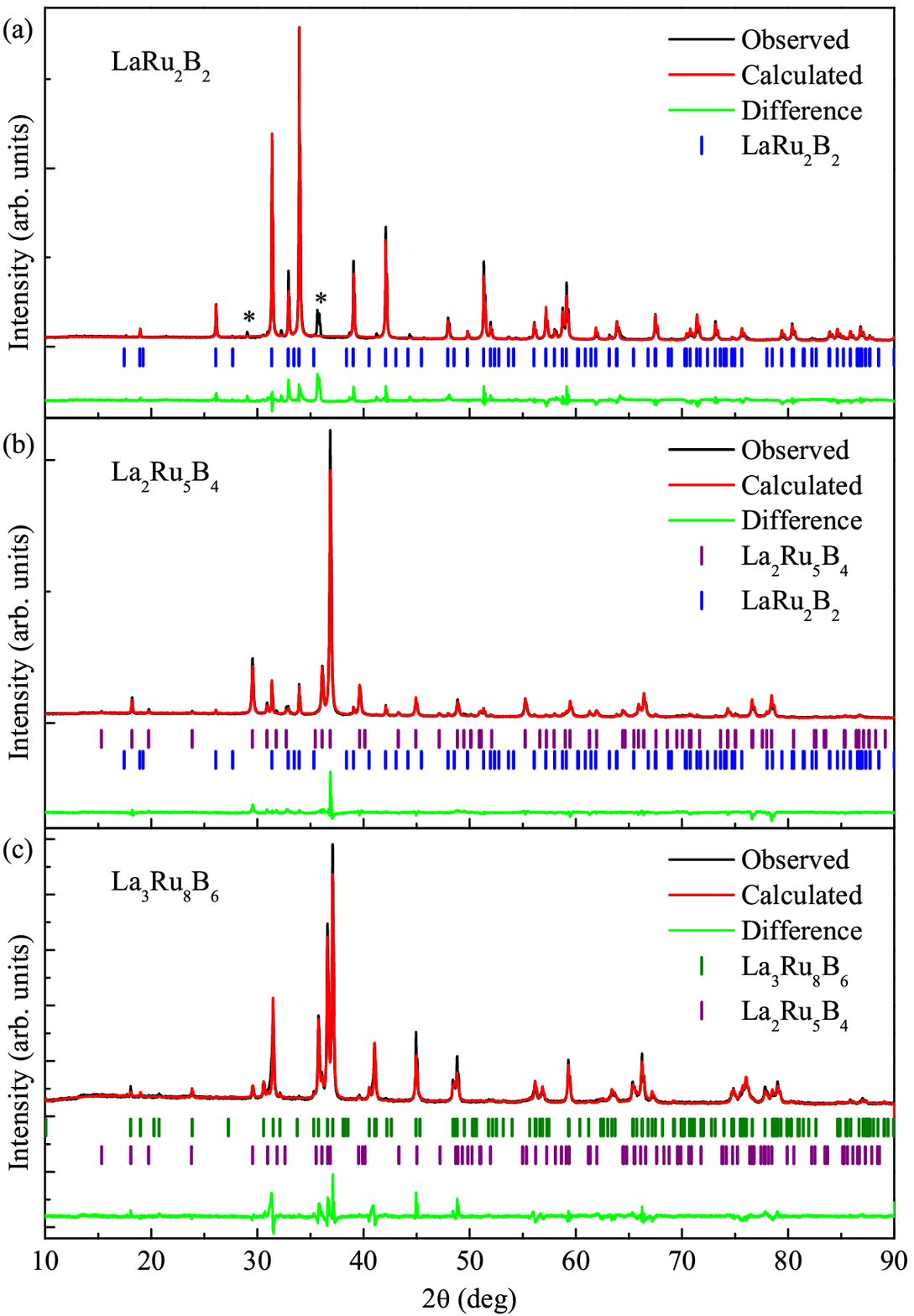}} \vspace*{-0.3cm}
\caption{Powder XRD patterns of (a) LaRu$_{2}$B$_{2}$, (b) La$_{2}$Ru$_{5}$B$_{4}$, and (c) La$_{3}$Ru$_{8}$B$_{6}$.}
\end{figure}

Figure 2 shows the powder X-ray diffraction (XRD) patterns of La$_{n}$Ru$_{3n-1}$B$_{2n}$ ($n=$ 1 - 3). For LaRu$_{2}$B$_{2}$ (fig. 2(a)), there is some unidentified peaks due to the unknown impurities. For La$_{2}$Ru$_{5}$B$_{4}$ (n = 2) (fig. 2(b)), it is a newly synthesized compound. The XRD pattern can be fitted well if using the structure of Sr$_{2}$Rh$_{5}$B$_{4}$ as the initial structure, indicating that La$_{2}$Ru$_{5}$B$_{4}$ is isostructrual to Sr$_{2}$Rh$_{5}$B$_{4}$. The fit is improved when the second phase LaRu$_{2}$B$_{2}$ is included. On the other hand, for La$_{3}$Ru$_{8}$B$_{6}$ (Fig. 2(c)), the XRD pattern can also be fitted very well by using two-phase fitting (La$_{3}$Ru$_{8}$B$_{6}$ (major) and La$_{2}$Ru$_{5}$B$_{4}$ (minor)). It is difficult to synthesize single-phase samples and the second phases always exist. This situation has also been pointed out in the literature when preparing A$_{n}$M$_{3n-1}$B$_{2n}$ (A = La, Y, Ca, and Sr, M = Ru, Os, and Rh) \cite{Takeya1,Horvath,Sologub1}. The existence of minor second phase with lower $n$ suggests that these homologous compounds might compete each other in the phase space. The fitted lattice parameters are listed in Table 1. The lattice parameters of LaRu$_{2}$B$_{2}$ and La$_{3}$Ru$_{8}$B$_{6}$ are consistent with previously reported values \cite{Horvath,Sologub1}.

\begin{table}[tbp] \centering
\caption{Space group and lattice parameters of La$_{n}$Ru$_{3n-1}$B$_{2n}$.}
\begin{tabular}{ccccc}
\hline\hline
Formula&S.G.&$a$ (\AA)&$b$ (\AA)&$c$ (\AA)\\
\hline
LaRu$_{2}$B$_{2}$&F222&6.4401(1)&9.2222(2)&10.1574(2)\\
La$_{2}$Ru$_{5}$B$_{4}$&Fmmm&5.6259(2)&9.7464(2)&11.5620(2)\\
La$_{3}$Ru$_{8}$B$_{6}$&Fmmm&5.5678(2)&9.8106(3)&17.5181(6)\\
\hline\hline
\end{tabular}
\label{TableKey}
\end{table}

\begin{figure}[tbp]
\centerline{\includegraphics[scale=0.65]{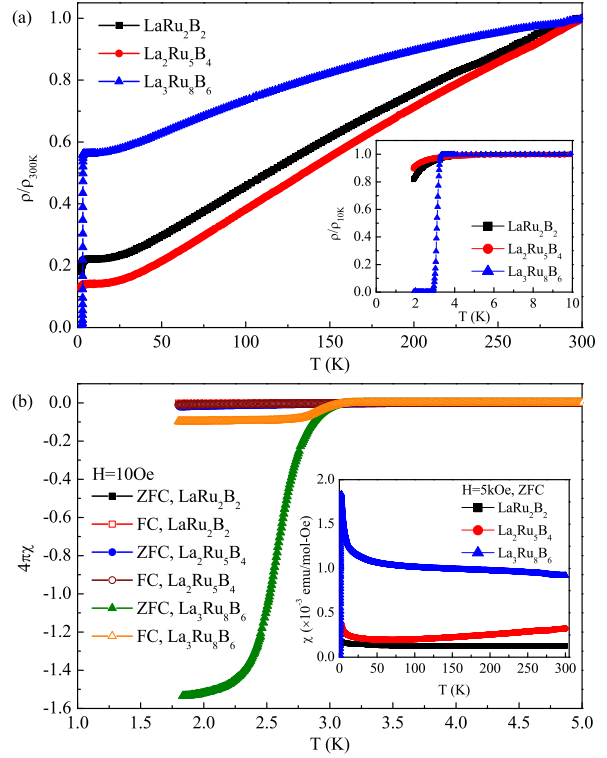}} \vspace*{-0.3cm}
\caption{(a) Temperature dependence of the normalized resistivity $\rho/\rho_{300K}(T)$ at zero field for La$_{n}$Ru$_{3n-1}$B$_{2n}$ ($n=$ 1 - 3). Inset: enlarged resistivity curves below 10 K. The resistivity are normalized to the values at 10 K. (b) Temperature dependence of $dc$ magnetic susceptibility $\chi(T)$ of La$_{n}$Ru$_{3n-1}$B$_{2n}$ ($n=$ 1 - 3) in the low temperature region and $H=$ 10 Oe with zero-field-cooling (ZFC) and field-cooling (FC) modes. Insert: $\chi(T)$ curves at $H=$ 5 kOe with ZFC mode between 1.8 and 300 K.}
\end{figure}

As shown in the main panel of fig. 3(a), all of samples exhibit metallic behaviors in the whole temperature range. In contrast to LaRu$_{2}$B$_{2}$ and La$_{2}$Ru$_{5}$B$_{4}$, the normalized resistivity of La$_{3}$Ru$_{8}$B$_{6}$ shows the saturation trend at high temperature. The saturation of $\rho(T)$ could be related to the Ioffe-Regel limit \cite{Ioffe}, i.e., the charge carrier mean free path is comparable to the interatomic spacing. Moreover, the residual resistivity ratio (RRR) of La$_{3}$Ru$_{8}$B$_{6}$ ($\sim$ 1.8) is also smaller than those ($\sim$ 4.6 and $\sim$ 7.1) in LaRu$_{2}$B$_{2}$ and La$_{2}$Ru$_{5}$B$_{4}$.  These results imply that La$_{3}$Ru$_{8}$B$_{6}$ might have shorter mean free path when compared to other two compounds. At low temperature, La$_{3}$Ru$_{8}$B$_{6}$ undergoes a relatively sharp superconducting transition at $T_{c,onset}$ = 3.31 K with transition width $\Delta T_{c}$ = 0.37 K when the samples with $n=$ 1 and 2 are still normal metals down to the lowest measuring temperature (inset of fig. 3(a)). There are slight drops at around 4 K in normalized resistivity curves for the samples with $n=$ 1 and 2. It could be due to the traces of superconducting impurities, such as LaRu$_{2}$ ($T_{c}=$ 4.1 K) \cite{Hillenbrand}, which could not be detected from XRD patterns.

The large diamagnetic signal of La$_{3}$Ru$_{8}$B$_{6}$ at $T=$ 1.8 K confirms the bulk superconductivity in La$_{3}$Ru$_{8}$B$_{6}$ (fig. 3(b)). The superconducting volume fraction (SVF) at 1.8 K is larger than 1 because of the uncorrected demagnetization factor of sample. The superconducting transition temperature ($T_{c}=$ 3.15 K) is consistent with that obtained from transport measurement. The $T_{c}$ of La$_{3}$Ru$_{8}$B$_{6}$ is close to those in isostructural Ca$_{3}$Rh$_{8}$B$_{6}$ ($T_{c}=$ 4 K) and Sr$_{3}$Rh$_{8}$B$_{6}$ ($T_{c}=$ 3.4 K) \cite{Takeya1}. The slightly wide superconducting transition width ($\sim$ 0.5 K) in $\chi(T)$ curve should be due to the existence of second phase. Moreover, the small SVF ($\sim$ 10\%) estimated from the FC curve indicates the rather strong vortex pinning effects in this compound. On the other hand, LaRu$_{2}$B$_{2}$ and La$_{2}$Ru$_{5}$B$_{4}$ do not show significant diagenetic signals in $\chi(T)$ curves at $T>$ 1.8 K, indicating that the absence of bulk superconductivity in these compounds. The very small SVF ($<$ 1.5\%) at 1.8 K originates from the impurities.

The $\chi(T)$ curves at high temperature are almost temperature-independent (inset of fig. 3(b)), suggesting the Pauli paramagnetism in these compounds. It is consistent with the metallic behavior shown in fig. 3(a). The rapid increase of $\chi(T)$ at low temperature could be due to the paramagnetic impurities. It should be noted that there is an upturn in La$_{2}$Ru$_{5}$B$_{4}$ at high temperature and the reason of behavior is not clear. If neglecting Van Vleck paramagnetism $\chi_{VV}$, the different contribution to the intrinsic $\chi$ at high temperate are $\chi=\chi_{core}+\chi_{L}+\chi_{P}$, where the $\chi_{core}$ is the isotropic diamagnetic susceptibility of localized core electrons, the $\chi_{L}$ is the generally isotropic Landau diamagnetic susceptibility of the conduction carriers, and the $\chi_{P}$ is the Pauli paramagnetic susceptibility of the conduction carriers. The $\chi_{core}$ are estimated using atomic diamagnetic susceptibilities \cite{Lawrence}, which give $\chi_{core}=$ -1.97, -4.37 and -5.58$\times$10$^{-4}$ emu/mol for LaRu$_{2}$B$_{2}$, La$_{2}$Ru$_{5}$B$_{4}$ and La$_{3}$Ru$_{8}$B$_{6}$, respectively. Assuming $\chi_{L}$ is not enhanced by the electron-phonon interaction, it gives $\chi_{L}\sim-\frac{1}{3}\chi_{P}$. Using the values of $\chi$ at 300 K, we obtain $\chi_{P}=$ 4.86, 11.4 and 22.14$\times$10$^{-4}$ emu/mol for LaRu$_{2}$B$_{2}$, La$_{2}$Ru$_{5}$B$_{4}$ and La$_{3}$Ru$_{8}$B$_{6}$, respectively. Such large $\chi_{P}$ values strongly imply there is an enhanced Pauli paramagnetism. The $\chi_{P}$ is related to $N(E_{F})$ by $\chi_{P}=\mu_{0}\mu_{B}^{2}N(E_{F})$, where $\mu_{0}$ is vacuum permeability, $\mu_{B}$ is Bohr magneton. We obtain $N(E_{F})=$ 15.03, 35.26 and 68.49 states/eV-f.u. for LaRu$_{2}$B$_{2}$, La$_{2}$Ru$_{5}$B$_{4}$ and La$_{3}$Ru$_{8}$B$_{6}$, respectively. As discussed below, these values are much larger than calculated values, implying the strong electron-electron interaction.

\begin{figure}[tbp]
\centerline{\includegraphics[scale=0.17]{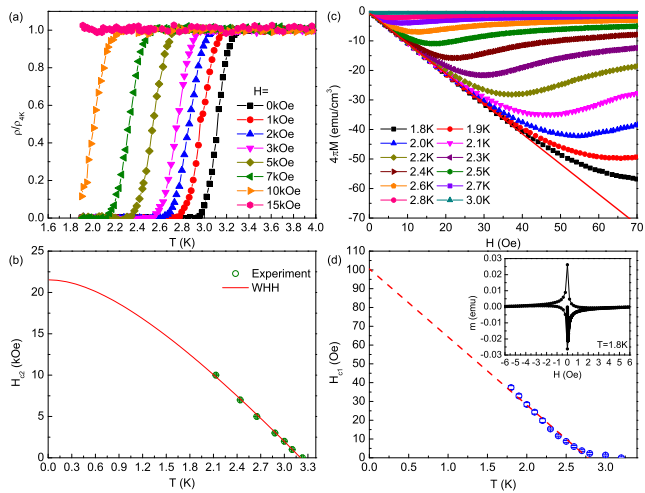}} \vspace*{-0.3cm}
\caption{(a) Temperature dependence of normalized resistivity $\rho/\rho_{4K}(T)$ at various magnetic fields for La$_{3}$Ru$_{8}$B$_{6}$. (b) Temperature dependence of $H_{c2}(T)$. Solid line shows the fitting result using WHH formula. (c) Low field parts of $M(H)$ at various temperature with demagnetization correction. The red solid lines are the "Meissner line" as discussed in the text. (d) Temperature dependence of $H_{c1}$. The red solid line is the linear fitting result. Inset: Magnetization hysteresis loops at $T=$ 1.8 K.}
\end{figure}

As shown in fig. 4(a), with increasing magnetic fields, the superconducting transition gradually shifts to the lower temperature but the transition width is almost unchanged. These observations indicate that the vortex-liquid state region is very narrow in La$_{3}$Ru$_{8}$B$_{6}$. At H = 15 kOe, the superconducting transition can not be observed above 1.9 K. Figure 4(b) shows the upper critical field $H_{c2}(T)$ of La$_{3}$Ru$_{8}$B$_{6}$ corresponding to the temperatures where the normalized resistivity drops to 90\%. Using the conventional one-band Werthamer-Helfand-Hohenberg (WHH) theory without considering Pauli spin paramagnetism effect and spin-orbit interaction \cite{Werthamer}, the $H_{c2}(T)$ can be fitted well when setting $T_{c}(H=0)$ and the slope $-dH_{c2}/dT|_{T=T_{c,H=0}}$ as free parameters (Fig. 4(b), red solid line). The obtained $-dH_{c2}/dT|_{T=T_{c,H=0}}$ = 9.7(2) kOe/K and $T_{c}$ = 3.21(1) K, consistent with the $T_{c,H=0}$ = 3.23(1) K. The estimated $H_{c2}(0)$ is 21.5(5) kOe. From the $H_{c2}(0)$ zero-temperature coherence length $\xi (0)$ can be estimated with Ginzburg-Landau formula $H_{c2}(0)$ = $\Phi _{0}/[2\pi \xi^{2}(0)]$, where $\Phi _{0}$ = 2.07$\times $10$^{-15}$ Wb. The derived $\xi (0)$ is 12.4(1) nm. The Pauli limiting field $H_{p}(0)$ = 18.4$T_{c}$ $\sim$ 61 kOe \cite{Clogston}, which is $\sim$ 3 times larger than that obtained from simplified WHH formula, therefore the orbital effect should be the dominant pair-breaking mechanism in La$_{3}$Ru$_{8}$B$_{6}$. This is similar to the case in (Ca,Sr)$_{3}$Rh$_{8}$B$_{6}$ compounds \cite{Takeya1}.

The shape of $m(H)$ loop at 1.8 K (inset of fig. 4(d)) points that La$_{3}$Ru$_{8}$B$_{6}$ is a typical type-II superconductor. Figure 4(c) shows the initial $M(H)$ curves at the low-field region. Linear region at low fields describes the Meissner shielding effects (¡°Meissner line¡±). The value of $H_{c1}^{\ast}$ at which the field starts to penetrate into the sample can be determined by examining the point of deviation from the Meissner line on the initial slope of the magnetization curve. But the $H_{c1}^{\ast}$ is not the same as the real lower critical field, due to the geometric effect. The $H_{c1}$ can be deduced from the first penetration field $H_{c1}^{\ast}$, assuming that the magnetization $M=-H _{c1}$ when the first vortex enters into the sample. Thus $H$ has been rescaled to $H=H_{a} - NM$ and $H_{c1} = H_{c1}^{\ast }/(1 - N)$ where $N$ is the demagnetization factor and $H_{a}$ is the external field \cite{Fossheim}. We estimate demagnetization factors to be 0.30 by using $H_{c1}$ = $H_{c1}^{\ast}$ /tanh($\sqrt{0.36b/a}$), where $a$ and $b$ are width and thickness of a plate-like superconductor \cite{Brandt}. After considering the demagnetization factors, the obtained slope of the linear fitting at the lowest temperature of our measurements $T=$ 1.8 K are -1.024(5), very close to -1 ($4\pi M=-H$). $H_{c1}$ is determined as the point deviating from linearity based on the criterion $\Delta m=3.5\times 10^{-4}$ emu, which is shown in fig. 4(d). The linear extrapolation of $H_{c1}(T)$ to $T=$ 0 K derives the $H_{c1}$(0) to be 100(4) Oe. Based on the values of $H_{c1}(0)$ and $H_{c2}(0)$, the Ginzburg-Landau (GL) parameter $\kappa$ can be obtained from $H_{c2}(0)/H_{c1}(0)$ = $2\kappa ^{2}/(ln\kappa +0.08)$. And thermodynamic critical field $H_{c}(0)$ can be obtained from $H_{c}(0)$ = $H_{c2}(0)/[\sqrt{2}\kappa(0)]$. The GL penetration length $\lambda$(0) can be evaluated using $\kappa$(0) = $\lambda$(0)/$\xi$(0) \cite{Plakida}. All of obtained parameters are listed in Table 2. The obtained $\kappa$ (= 17.8(6)) confirms that La$_{3}$Ru$_{8}$B$_{6}$ is the type-II superconductor, consistent with the feature of $m(H)$ loop below $T_{c}$.

\begin{table}[tbp] \centering
\caption{Superconducting parameters of La$_{3}$Ru$_{8}$B$_{6}$.}
\begin{tabular}{ccccccccccccccccc}
\hline\hline
&$T_{c}$ (K)&&&&&&3.2\\
&$H_{c2}(0)$ (kOe)&&&&&&21.5(5)\\
&$H_{c1}(0)$ (Oe)&&&&&&100(4)\\
&$H_{c}(0)$ (Oe)&&&&&&852(49)\\
&$\xi(0)$ (nm)&&&&&&12.4(1)\\
&$\lambda(0)$ (nm)&&&&&&221(9)\\
&$\kappa_{GL}$&&&&&&17.8(6)\\
&$\Delta{C}/\gamma{T_{c}}$&&&&&&0.89(1)\\
&$\lambda_{e-ph}$&&&&&&0.56\\
\hline\hline
\end{tabular}
\label{TableKey}
\end{table}

\begin{figure}[tbp]
\centerline{\includegraphics[scale=0.7]{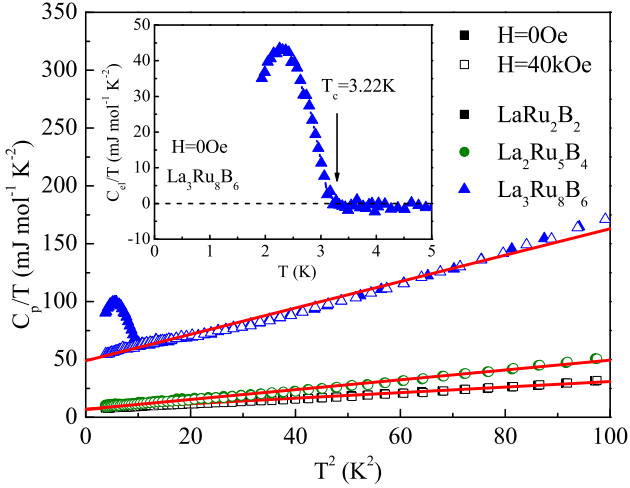}} \vspace*{-0.3cm}
\caption{(a) Low temperature specific heat  $C_{p}/T$ $vs.$ $T^{2}$ from 1.92 K to 10 K at $H=$ 0 and 40 kOe. The solid lines show the fits (see text). Inset: Temperature dependence of the electronic specific heat plotted as $C_{el}/T$ vs. $T$ at zero field for La$_{3}$Ru$_{8}$B$_{6}$.}
\end{figure}

The main panel of fig. 5 shows the temperature dependence of the specific
heat of La$_{n}$Ru$_{3n-1}$B$_{2n}$ ($n=$ 1 - 3) below 10 K at $H=$ 0 and 40 kOe. For La$_{3}$Ru$_{8}$B$_{6}$, a jump at $T=$ 3.2 K can be clearly seen at zero field (inset of fig. 5), consistent with the $T_{c}$ determined from transport and susceptibility measurements. The specific heat jump confirms the bulk superconductivity of La$_{3}$Ru$_{8}$B$_{6}$. At $H=$ 40 kOe, the superconducting transition is suppressed below 1.9 K. On the other hand, LaRu$_{2}$B$_{2}$ and La$_{2}$Ru$_{5}$B$_{4}$ do not exhibit the jump in specific heat curves, indicating the absent of bulk superconductivity. In order to obtain the normal state electronic specific heat coefficient $\gamma $ and Debye temperatures $\Theta _{D}$, the low temperature specific heat at $H=$ 40 kOe for all of compounds are fitted using $C_{p}/T$ = $\gamma +\beta T^{2}$ (red lines in fig. 5). The obtained $\gamma$ and $\Theta _{D}$ using $\Theta_{D}$ = $(12\pi ^{4}NR/5\beta )^{1/3}$ where N is the number of atoms per formula unit and R is the gas constant, are listed in Table 3. It can be seen that the $\Theta _{D}$ for these compounds are close each other, and also similar to the values in (Ca,Sr)$_{3}$Rh$_{8}$B$_{6}$. However, the values of $\gamma$ are much larger than those in (Ca,Sr)$_{n}$Rh$_{3n-1}$B$_{2n}$ \cite{Takeya1}. Moreover, the value $\gamma$ in La$_{3}$Ru$_{8}$B$_{6}$ is significantly larger than other two compounds, similar to the case in (Ca,Sr)$_{n}$Rh$_{3n-1}$B$_{2n}$ \cite{Takeya1}. The $\gamma$ is related to $N(E_{F})$ by $\gamma=\frac{\pi^{2}}{3}k_{B}^{2}N(E_{F})$, where $k_{B}$ is Boltzmann constant. We obtain $N(E_{F})$ is 2.88(4), 2.93(8) and 20.8(2) states/eV-f.u. for LaRu$_{2}$B$_{2}$, La$_{2}$Ru$_{5}$B$_{4}$ and La$_{3}$Ru$_{8}$B$_{6}$, respectively. These values are larger than the calculated values as shown below, suggesting the remarkable electron-phonon and/or electron-electron interaction.

 According to the McMillan formula for electron-phonon mediated superconductivity \cite{McMillan}, the electron-phonon coupling constant $\lambda _{e-ph}$ in La$_{3}$Ru$_{8}$B$_{6}$ can be determined by

\begin{equation}
\lambda_{e-ph}=\frac{\mu ^{\ast }\ln(1.45T_{c}/\Theta _{D})-1.04}{%
1.04+\ln(1.45T_{c}/\Theta _{D})(1-0.62\mu ^{\ast })}
\end{equation}

where $\mu^{\ast }\approx $ 0.13 is the common value for Coulomb
pseudopotential. By using $T_{c}$ = 3.2 K and $\Theta _{D}$ = 307 K, we obtain
$\lambda _{e-ph}=$ 0.56, indicating an intermediately coupled superconductor. The
electronic specific heat part $C_{el}$ is obtained by $C_{el}=C_{p}(H=0)-C_{p}(H=$ 40 kOe$)$ (inset of fig. 5). The specific heat jump at $T_{c}$, $\Delta $C$_{el}$/$\gamma T_{c}=$ 0.89(1), is smaller than the weak coupling value 1.43 \cite{McMillan}, which could be partially due to the existence of small amount of non-superconducting impurities in the sample. Another reason could be the multi-band effect which has been intensively studied on MgB$_{2}$.\cite{Bouquet} Moreover, the gap anisotropy could also lead to the small $\Delta $C$_{el}$/$\gamma T_{c}$, such as in the case of Nb$_{3}$Se$_{4}$ \cite{Okamoto}. On the other hand, the anomaly large $\gamma$ could also make decrease the value of $\Delta $C$_{el}$/$\gamma T_{c}$. It is similar to the heavy fermion superconductors, such as CeIrIn$_{5}$, CeCu$_{2}$Si$_{2}$ and UPt$_{3}$ \cite{Petrovic, Grewe}.

\begin{figure}[tbp]
\centerline{\includegraphics[scale=0.6]{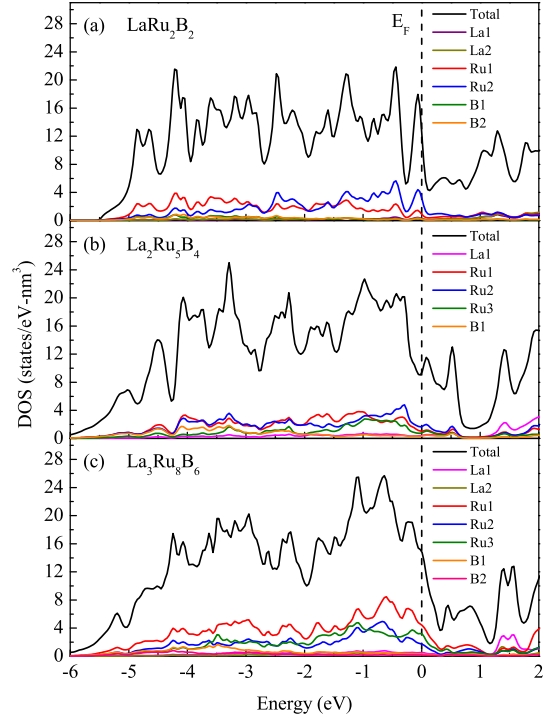}} \vspace*{-0.3cm}
\caption{The calculated total and partial DOS curves for (a) LaRu$_{2}$B$_{2}$, (b) La$_{2}$Ru$_{5}$B$_{4}$, and (c) La$_{3}$Ru$_{8}$B$_{6}$ (in units of states/eV-nm$^{3}$).}
\end{figure}

\begin{table*}[tbp] \centering
\caption{Debye temperature $\Theta_{D}$ (in unit of K), normal state electronic specific heat coefficient $\gamma$ (in units of mJ/mol-K$^{2}$), derived $N(E_{F})_{T}$ from $\gamma$ (in units of states/eV-f.u.), Pauli paramagnetic susceptibility $\chi_{P}$ (in units of emu/mol-Oe), derived $N(E_{F})_{M}$ from $\chi_{P}$ (in units of states/eV-f.u.), calucalted $N(E_{F})_{cal}$ (in units of states/eV-f.u.), calculated $N(E_{F})^{V}_{cal}$ (in units of states/eV-nm$^{3}$) and Wilson ratio $R_{W}$ of La$_{n}$Ru$_{3n-1}$B$_{2n}$ ($n=1-3$).}
\begin{tabular}{ccccccccccccccccc}
\hline\hline
Formula&$\Theta_{D}$&&$\gamma$&&$N(E_{F})_{T}$&&$\chi_{P}$&&$N(E_{F}$)$_{M}$&&$N(E_{F})_{cal}$&&$N(E_{F})^{V}_{cal}$&&$R_{W}$\\
\hline
LaRu$_{2}$B$_{2}$&343(1)&&6.8(1)&&2.88(4)&&4.86&&15.03&&0.95&&12.55&&5.21(8)\\
La$_{2}$Ru$_{5}$B$_{4}$&369(1)&&6.9(2)&&2.93(8)&&11.40&&35.26&&1.51&&9.53&&12.1(3)\\
La$_{3}$Ru$_{8}$B$_{6}$&307(1)&&48.9(5)&&20.8(2)&&22.14&&68.49&&3.54&&14.78&&3.30(3)\\
\hline\hline
\end{tabular}
\label{TableKey}
\end{table*}

Figure 6 shows the calculated total and partial DOS curves of La$_{n}$Ru$_{3n-1}$B$_{n2}$ ($n=$ 1 - 3) near $E_{F}$. It should be noted that the units of DOS is changed to states/eV-nm$^{3}$ in order to compare the DOSs in these homologous compounds. The finite DOSs at $E_{F}$ indicate that they are metal, consistent with the results of transport measurement. The Ru atoms have the main contribution to the DOS near $E_{F}$ for all three compounds, similar to (Ca,Sr)$_{n}$Rh$_{3n-1}$B$_{2n}$ \cite{Takeya1}. Moreover, the $N(E_{F})_{cal}$ values (in units of states/eV-f.u.) (listed in Table 3) are also closed to the values in the latter \cite{Takeya1}. On the other hand, the $N(E_{F})^{V}_{cal}$ (in units of states/eV-nm$^{3}$) of La$_{3}$Ru$_{8}$B$_{6}$ (listed in Table 3) is largest among these compounds. Thus, the origin of superconductivity in La$_{3}$Ru$_{8}$B$_{6}$ could be partially ascribed to the high $N(E_{F})$, similar to (Ca,Sr)$_{3}$Rh$_{8}$B$_{6}$ \cite{Takeya1}. But the dependence of $T_{c}$ on $n$ is not monotonic, i.e., the larger $n$ does not mean the higher $T_{c}$, because La$_{1-\delta}$Ru$_{3}$B$_{2}$ ($\delta \sim$ 0.1, $n=\infty$) does not show superconducting transition when $T>$ 1.2 K \cite{Ku}. It suggest that there might be an optimal $n$ for highest $T_{c}$ and other factors, such as the strength of electron - phonon coupling and detailed phonon spectrum related to the detailed structure, could also have some influences on the $T_{c}$.

From Table 3, it can be seen that $N(E_{F})_{cal}$ is much smaller than the derived values from susceptibility and specific heat measurements, suggesting that there are strong electron-phonon and/or electron-electron interactions. In order to estimate the strength of the electron-electron correlation effect in these materials, we calculate the Wilson ratio ($R_{W}=\frac{\pi^2k_{B}^2}{3\mu_{0}\mu_{B}^2}\frac{\chi_{P}}{\gamma}$), which measures the relative enhancements of the spin susceptibility and electronic specific heat \cite{Wilson}. For a non-interacting Fermi liquid, $R_{W}$ is expected to be close to 1. The rather large $R_{W}$ (Table 3) indicates the strong electron-electron correlation in these materials. Such large $R_{W}$ has been found in the heavy fermi liquids ($R_{W}=$ 1 - 6) \cite{Delong}, and in the system with a magnetic instability or strong exchange enhanced paramagnetic state \cite{Julian}. Moreover, interestingly, the strong electron-electron correlation has also been found in LaRu$_{3}$Si$_{2}$ superconductor with $T_{c}=$ 7.8 K \cite{Li}. It has the CeCo$_{3}$B$_{3}$-type structure closely related to the structures we discussed in this work.

\section{Conclusion}
In summary, we studied the physical properties of homologous series La$_{n}$Ru$_{3n-1}$B$_{2n}$ ($n=1-3$) compounds. Among these compounds, La$_{2}$Ru$_{5}$B$_{4}$ is a new member with $n=$ 2. LaRu$_{2}$B$_{2}$ and La$_{2}$Ru$_{5}$B$_{4}$ show normal metallic behaviors down to 1.8 K. In contrast, La$_{3}$Ru$_{8}$B$_{6}$ is an intermediately coupled BCS superconductor with $T_{c}\sim$ 3.2 K. The experimental and theoretical calculation results suggest that the emergence of superconductivity in La$_{3}$Ru$_{8}$B$_{6}$ could be due to the rather large $N(E_{F})$, similar to (Ca, Sr)$_{3}$Rh$_{8}$B$_{6}$. On the other hand, all of these compounds show strong electron-electron correlation. Because of the preference of superconductivity to the structure of A$_{3}$M$_{8}$B$_{6}$ (A = Ca, Sr, La, and Y; M = Ru, Os, and Rh), it is of interest to explore other new superconductors with this structure.

\acknowledgments
This work was supported by the Funding Program for World-Leading Innovative R\&D on Science and Technology (FIRST), Japan.

\end{document}